\documentclass[aip,
nofootinbib,
rsi,%
reprint,
twocolumn,
longbibliography,
author-numerical%
]{revtex4-2}

\usepackage{amsmath,amssymb,graphicx}
\usepackage[usenames]{color}

\usepackage{setspace}
\usepackage{lineno}
\usepackage{comment}

\newcommand{\pdr}[2]{\dfrac{\partial {#1}}{\partial {#2}}}
\newcommand{\pddr}[2]{\dfrac{\partial^2 {#1}}{\partial {#2}^2}}

\newcommand{\tx}{\tilde{x}}

\newcommand{\tj}{\tilde{j}}

\newcommand{\tR}{\tilde{R}}

\newcommand{\teta}{\tilde{\eta}}

\newcommand{\tZ}{\tilde{Z}}

\newcommand{\tD}{\tilde{D}}
\newcommand{\tc}{\tilde{c}}

\newcommand{\lexp}[1]{\exp\left(#1\right)}

\newcommand{\cref}{c_{ref}}

\newcommand{\lcat}{l_t}

\newcommand{\etal}{et{} al.{} }

\newcommand{\Dox}{D_{ox}}
\newcommand{\tDox}{\tD_{ox}}

\newcommand{\tit}{\tilde{t}}
\newcommand{\Cdl}{C_{dl}}

\newcommand{\ri}{{\rm i}}
\newcommand{\tom}{\tilde{\omega}}

\renewcommand{\Im}[1]{\operatorname{Im}\left(#1\right)}

\begin{document}

\sf

\title{Performance of a PEM fuel cell
   cathode catalyst layer under oscillating  potential and oxygen supply}

\author{Andrei Kulikovsky}
\email{A.Kulikovsky@fz-juelich.de}

\affiliation{Forschungszentrum J\"ulich GmbH           \\
    Theory and Computation of Energy Materials (IEK--13)   \\
    Institute of Energy and Climate Research,              \\
    D--52425 J\"ulich, Germany
}

\altaffiliation[Also at: ]{Lomonosov Moscow State University,
    Research Computing Center, 119991 Moscow, Russia}

\date{\today}

\begin{abstract}
A model for impedance of a PEM fuel cell cathode catalyst layer under
simultaneous application of potential and oxygen concentration
perturbations is developed and solved. The resulting expression
demonstrates dramatic lowering of the layer impedance under increase
in the amplitude of the oxygen concentration perturbation.
In--phase oscillations of the overpotential and oxygen concentration
lead to formation of a fully transparent to oxygen sub--layer.
This sub--layer works as an ideal non polarizable electrode,
which strongly reduces the system impedance.
\end{abstract}

\keywords{PEM fuel cell, catalyst layer, impedance, modeling}

\maketitle

\section{Introduction}

Electrochemical impedance spectroscopy (EIS) has proven to be a unique non--destructive
and non--invasive tool for fuel cells characterization~\cite{Lasia_book_14}.
In its classic variant, EIS implies application of a small--amplitude harmonic
perturbation of the cell current or potential and measuring the response
of the cell potential or current, respectively. In recent years, there has been
interest in alternative techniques based on application
of pressure~(Engebretsen \etal\cite{Brett_17},
Shirsath \etal\cite{Bessler_20}, Schiffer \etal\cite{Bessler_22},
Zhang \etal\cite{Zhang_22})
or oxygen concentration~(Sorrentino \etal\cite{Sorrentino_17,Sorrentino_19,Sorrentino_20})
perturbation to the cell and measuring the response of
electric variable (potential or current), keeping the second electric variable
constant.

Application of pressure oscillations at the cathode channel inlet
or outlet inevitably leads to flow velocity oscillations (FVO).
Kim \etal~\cite{Kim_08b} and Hwang \etal~\cite{Hwang_10}
reported experiments showing dramatic improvement of PEM fuel cell performance
under applied FVO. The effect of FVO on the cell performance was more pronounced
with lower static flow rates and with increasing the FVO amplitude~\cite{Kim_08b}.
In~\cite{Kim_08b,Hwang_10}, the effect has been attributed to improvement
of diffusive oxygen transport through the cell due to FVO.
Kulikovsky~\cite{Kulikovsky_20arXiv,Kulikovsky_21a} developed a simplified
analytical model for impedance of the cell subjected to simultaneous oscillations
of potential and air flow velocity. The model have shown reduction of the
cell static resistivity upon increase of the FVO amplitude.
Yet, however, due to system complexity,
the mechanism of cell performance improvement is not clear.

Below, a much simpler system (PEM fuel cell cathode catalyst layer) subjected to
oscillating in--phase potential and oxygen supply is considered. Analytical model
for the CCL impedance under these conditions is solved.
The result demonstrates the effect of impedance reduction due to oscillating oxygen supply.
In--phase oxygen concentration and overpotential oscillations make part of
the catalyst layer at the cathode catalyst layer (CCL)/gas diffusion layer (GDL) interface
fully transparent to oxygen, which leads
to dramatic decrease of the system impedance.

\section{Model}

\begin{figure}
    \begin{center}
        \includegraphics[scale=0.7]{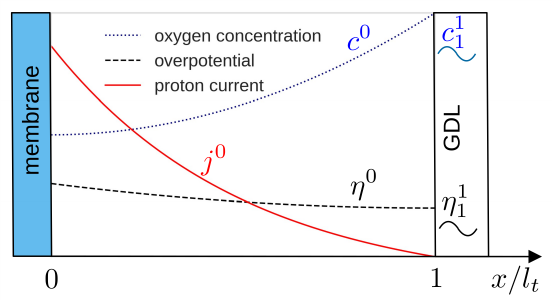} \\
        \caption{Schematic of the cathode catalyst layer
        and typical shapes of the oxygen concentration, proton current and
        overpotential through the layer. In--phase harmonic perturbations of the
        overpotential $\eta_1^1$ and oxygen concentration $c_1^1$ are applied
        at the CCL/GDL interface.
        }
        \label{fig:template}
    \end{center}
\end{figure}

Consider a problem for impedance of the cathode catalyst layer (CCL)
under oscillating potential and oxygen supply (Figure~\ref{fig:template}).
For simplicity, we will assume that the proton transport is fast.
The model is based on two equations: the proton charge conservation
\begin{equation}
    \Cdl\pdr{\eta}{t} + \pdr{j}{x} = - i_*\left(\dfrac{c}{\cref}\right)\lexp{\dfrac{\eta}{b}}
    \label{eq:etax}
\end{equation}
and the oxygen mass transport equation
\begin{equation}
    \pdr{c}{t} - \Dox\pddr{c}{x}
    = - \dfrac{i_*}{4F}\left(\dfrac{c}{\cref}\right)\lexp{\dfrac{\eta}{b}}.
    \label{eq:cox}
\end{equation}
Here,
$x$ is the distance through the CCL,
$\Cdl$ is the double layer capacitance,
$\eta$ is the positive by convention ORR overpotential,
$t$ is time,
$j$ is the local proton current density,
$x$ is the coordinate through the CCL,
$i_*$ is the ORR volumetric exchange current density,
$\cref$ is the reference oxygen concentration, and
$b$ is the ORR Tafel slope.
Introducing dimensionless variables
\begin{multline}
    \tx = \dfrac{x}{\lcat}, \quad \tit = \dfrac{t i_*}{\Cdl b},
    \quad \tc = \dfrac{c}{\cref}, \quad \teta = \dfrac{\eta}{b},
    \quad \tj = \dfrac{j}{i_*\lcat}, \\
    \tDox = \dfrac{4 F \Dox \cref }{i_* \lcat^2}, \quad
    \tZ = \dfrac{Z i_*\lcat}{b}, \quad  \tom = \dfrac{\omega \Cdl b}{i_*},
    \label{eq:dless}
\end{multline}
where
$\tom$ is the angular frequency of applied AC signal,
$\tZ$ is the CCL impedance,
Eqs.\eqref{eq:etax}, \eqref{eq:cox} take the form
\begin{align}
    &\pdr{\teta}{\tit} + \pdr{\tj}{\tx} = - \tc\exp\teta
    \label{eq:tetax} \\
    &\mu^2\pdr{\tc}{\tit} - \tDox\pddr{\tc}{\tx} = - \tc\exp\teta,
    \label{eq:tcox}
\end{align}
where
$\mu$ is the constant parameter
\begin{equation}
    \mu = \sqrt{\dfrac{4 F\cref}{\Cdl b}}.
    \label{eq:mu}
\end{equation}

Substituting  Fourier--transforms of the form
\begin{equation}
\begin{split}
   &\teta(\tit) = \teta^0 + \teta_1^1(\tom) \exp(\ri\tom\tit), \quad |\teta_1^1| \ll \teta^0 \\
   &\tj(\tx, \tit) = \tj^0(\tx) + \tj^1(\tx,\tom) \exp(\ri\tom\tit), \quad |\tj^1| \ll \tj^0 \\
   &\tc(\tx, \tit) = \tc^0(\tx) + \tc^1(\tx,\tom) \exp(\ri\tom\tit), \quad |\tc^1| \ll \tc^0,
\end{split}
\label{eq:F}
\end{equation}
into Eqs.\eqref{eq:tetax}, \eqref{eq:tcox},
neglecting terms with the perturbation products and subtracting the respective
static equations, we get a system of equations relating the perturbation
amplitudes $\teta_1^1(\tom)$, $\tj^1(\tx,\tom)$ and $\tc^1(\tx,\tom)$~\cite{Kulikovsky_16a}:
\begin{equation}
    \pdr{\tj^1}{\tx}
       = - \exp(\teta^0)\left(\tc^1 + \tc_1^0\teta_1^1 \right) - \ri\tom\teta_1^1,
       \quad  \tj^1(1)=0
    \label{eq:atetax}
\end{equation}
\begin{multline}
    \tDox\pddr{\tc^1}{\tx} = \exp(\teta^0) \left(\tc^1 + \tc_1^0\teta_1^1\right)
        + \ri\mu^2\tom\tc^1, \\
        \left.\pdr{\tc^1}{\tx}\right|_{\tx=0} = 0, \quad \tc^1(1) = k\teta_1^1
    \label{eq:atcx}
\end{multline}
where the superscripts 0 and 1 mark the static variables and
the small perturbation amplitudes, respectively,
$\tc_1^0$ is the static oxygen concentration at the CCL/GDL interface,
$k \geq 0$ is the constant model parameter,
$\teta_1^1$ is the amplitude of applied potential perturbation.

The boundary condition for Eq.\eqref{eq:atetax} means zero proton current at the CCL/GDL
interface. The left boundary condition for Eq.\eqref{eq:atcx} expresses zero oxygen flux
through the membrane. The feature of this problem is the
right boundary condition for Eq.\eqref{eq:atcx}, meaning external
control of oxygen concentration perturbation $\tc^1$ at the CCL/GDL interface:
$\tc^1(1)$ varies in--phase with the overpotential perturbation $\teta_1^1$.
Note that the perturbations of applied cell potential and overpotential
have different signs, assuming that the electron conductivity of the cell components
is much larger than the CCVL proton conductivity.

Due to fast proton transport, $\teta_1^1$ is nearly independent of $\tx$.
For simplicity we will also assume that the variation of
static oxygen concentration along $\tx$ is also small and we set $\tc^0 \simeq \tc_1^0$.
Introducing electric $Y$ and concentration $G$ admittances
\begin{equation}
    Y = \dfrac{\tj^1}{\teta_1^1}, \quad G = \dfrac{\tc^1}{\teta_1^1}
    \label{eq:YG}
\end{equation}
and taking into account the static polarization curve
\begin{equation}
    \tj_0 = \tc_1^0\exp\teta^0
    \label{eq:tvcc}
\end{equation}
we can rewrite the system \eqref{eq:atetax}, \eqref{eq:atcx} in terms of $Y$ and $G$:
\begin{equation}
    \pdr{Y}{\tx} = - \dfrac{\tj_0}{\tc_1^0}\left(G + \tc_1^0\right) - \ri\tom,
    \quad Y(1) = 0
    \label{eq:Yx}
\end{equation}
\begin{multline}
    \tDox\pddr{G}{\tx} = \dfrac{\tj_0}{\tc_1^0} \left(G + \tc_1^0\right) + \ri\mu^2\tom G, \\
     \left.\pdr{G}{\tx}\right|_{\tx=0} = 0, \quad G(1) = k.
    \label{eq:Gx}
\end{multline}
Eq.\eqref{eq:Gx} can be directly solved:
\begin{multline}
   G(\tx) = \dfrac{\tj_0}{\psi^2}\left(\dfrac{\cosh(\psi \tx)}
                                        {\cosh(\psi)} - 1\right)
     + \dfrac{k \cosh(\psi \tx)}{\cosh(\psi)}, \\
     \psi \equiv \sqrt{\dfrac{\tj_0}{\tc_1^0} + \ri\mu^2\tom}.
     \label{eq:Gsol}
\end{multline}
Using this solution and solving Eq.\eqref{eq:Yx}, for the CCL
impedance $\tZ = 1/Y|_{\tx=0}$ we get
\begin{equation}
    \tZ = \dfrac{\tZ_{RC}}
    {\left(\tj_0/\tc_1^0\right) \left(k \tZ_{RC} + \tj_0\right) \tZ_W
        + \left(\tj_0\mu^2 + \tZ_{RC}\right)\ri\tom},
    \label{eq:tZccl}
\end{equation}
where $\tZ_{RC}$ is the parallel $RC$--circuit impedance, and $\tZ_W$ is
the Warburg--like impedance:
\begin{equation}
    \tZ_{RC} = \dfrac{\tj_0}{\tc_1^0} + \ri\mu^2\tom, \quad
    \tZ_W = \dfrac{\tanh\left(\sqrt{\tZ_{RC}/\tDox}\right)}{\sqrt{\tZ_{RC}/\tDox}}
    \label{eq:tZrc}
\end{equation}

\begin{figure}
    \begin{center}
        \includegraphics[scale=0.45]{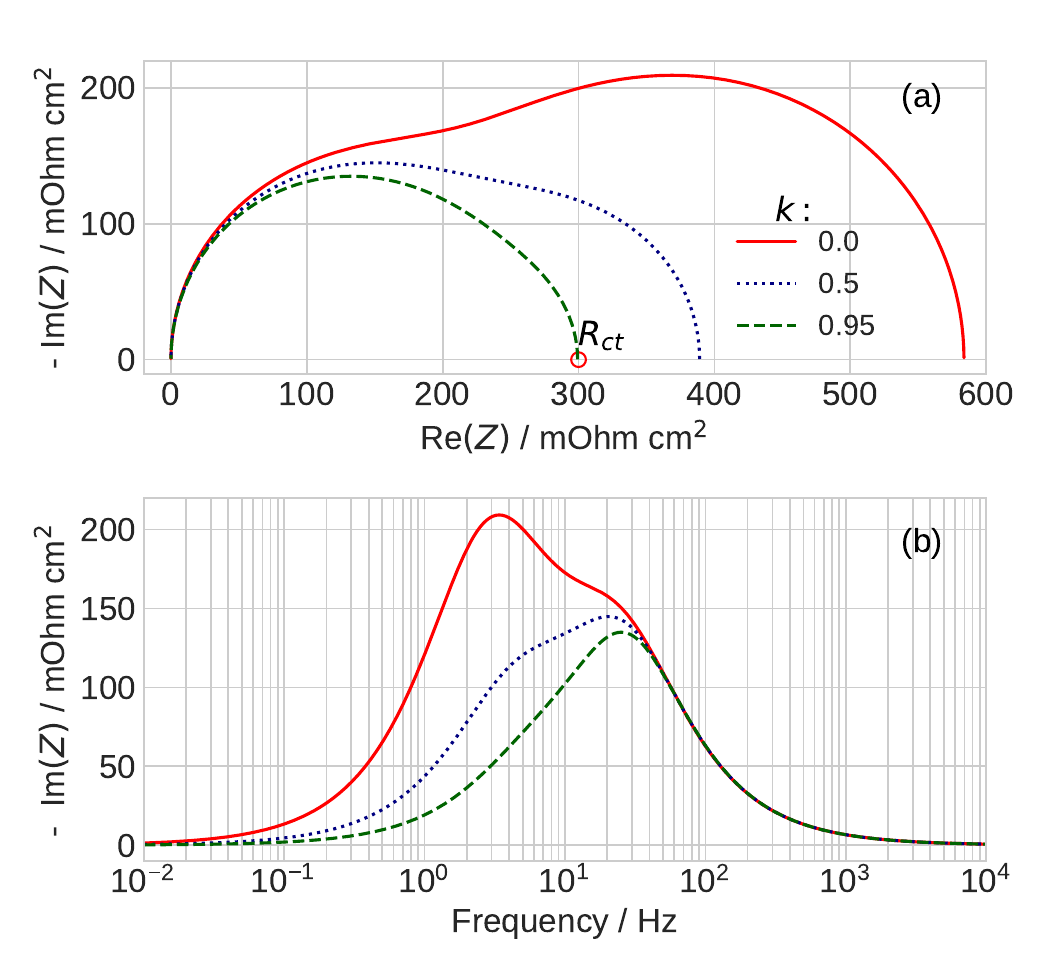} \\
        \caption{(a) Nyquist spectra of the CCL, Eq.\eqref{eq:tZccl},
            for the indicated values of parameter $k$. The other parameters are listed
            in Table~\ref{tab:parms}. The point $R_{ct}$ indicates the static charge--transfer
            resistivity $b/j_0$.
            (b) The frequency dependence of the impedances in (a).
        }
        \label{fig:Zccl}
    \end{center}
\end{figure}

CCL impedance $Z$, Eq.\eqref{eq:tZccl},
for the current density of 100 mA~cm$^{-2}$ and the
indicated values of parameter $k$
is shown in Figure~\ref{fig:Zccl}.
The other parameters are listed in Table~\ref{tab:parms}.
Note that to emphasize the effect,
the CCL oxygen diffusivity is taken to be very low, $\Dox = 10^{-5}$~cm$^2$~s$^{-1}$.

The Nyquist spectrum corresponding to $k=0$ (no concentration perturbation
at the CCL/GDL interface) consists
of two partly overlapping arcs, of which the left one is due to charge transfer and
the right one is due to oxygen transport (Figure~\ref{fig:Zccl}a). In Figure~\ref{fig:Zccl}b,
the left peak of $-\Im{Z}$ (red curve) corresponds to oxygen transport
and the right shoulder  is due to faradaic reaction.

The perturbation of oxygen concentration at the CCL/GDL interface
dramatically changes the Nyquist spectra. With the growth of $k$,
the arc due to oxygen transport strongly decreases (Figure~\ref{fig:Zccl}).
Calculation of Eq.\eqref{eq:tZccl} limit as $\tom \to 0$ gives
the CCL static resistivity
\begin{equation}
    \tR = \dfrac{\phi\coth(\phi)}{\tj_0\left(1 + k/\tc_1^0\right)},
    \quad \phi \equiv \sqrt{\dfrac{\tj_0}{\tc_1^0\tDox}}
     \label{eq:tRccl}
\end{equation}
which decreases as $k$ increases (Figure~\ref{fig:Zccl}a).

\begin{figure}
    \begin{center}
        \includegraphics[scale=0.45]{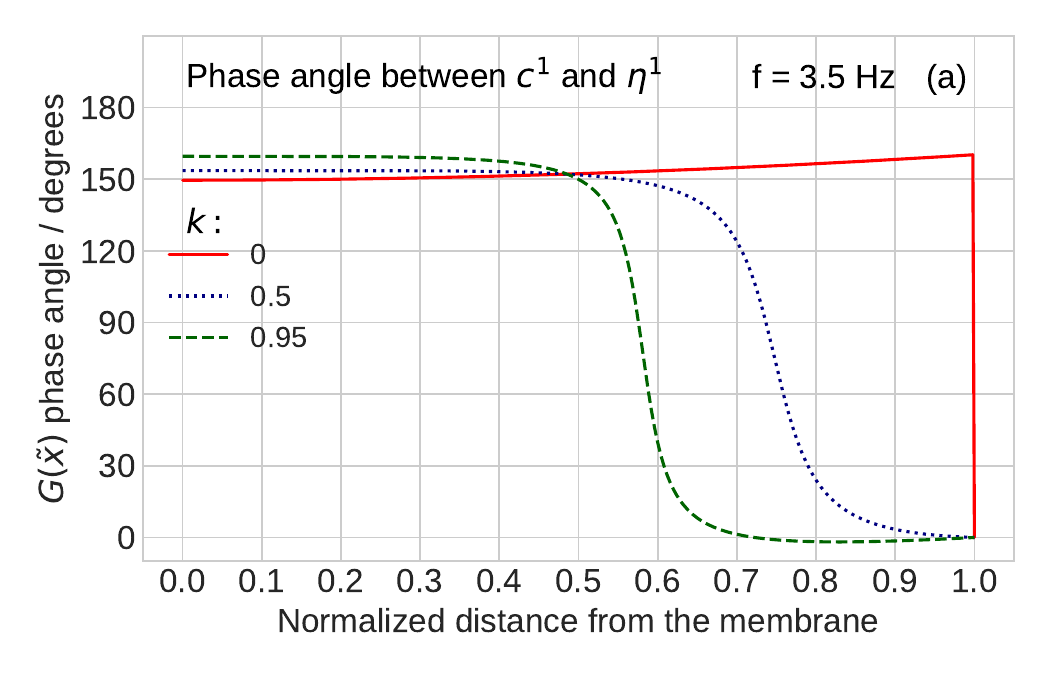} \\
        \includegraphics[scale=0.45]{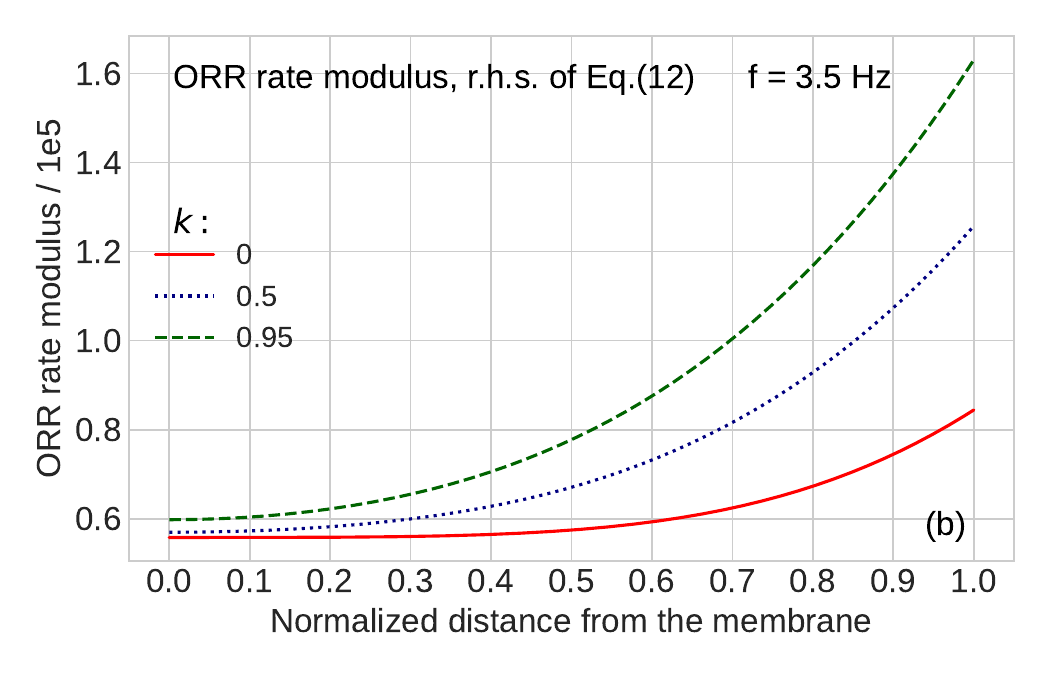} \\
        \caption{(a) The phase angle of $G = \tc^1/\teta_1^1$ (between
         oxygen concentration and overpotential) vs
         distance $\tx$ through the CCL. The frequency of 3.5 Hz corresponds
         to position of the transport peak in Figure~\ref{fig:Zccl}b.
         (b) The modulus of the right side of Eq.\eqref{eq:Yx},
          which is the transient ORR rate.
        }
        \label{fig:xshapes}
    \end{center}
\end{figure}

Figure~\ref{fig:xshapes}a shows what happens to the
$\tx$--shape of the phase shift between the
oxygen concentration  $\tc^1$ and overpotential  $\teta_1^1$
(the phase angle of the function $G(\tx) =\tc^1/\teta_1^1$)
as $k$ increases. At $k=0$, there is a large phase shift between
$\tc_1$ and $\teta_1^1$, excluding a single point at $\tx=1$, where
$\tc^1=0$. With $k=0.5$, a finite--thickness domain where $\tc^1$
and $\teta_1^1$ oscillate almost in--phase forms, and with $k=0.95$ this domain
of zero phase shift increases (Figure~\ref{fig:xshapes}).
Zero phase angle between $\teta_1^1$ and $\tc^1$ means no oxygen transport
losses in this domain.

Figure~\ref{fig:xshapes}b shows the modulus of the right side of Eq.\eqref{eq:Yx},
which is the modulus of the ORR rate perturbation. As can be seen,
in the domain transparent for oxygen, the ORR rate strongly increases.
Thus, with $k>0$, a relatively thin sub--layer at the CCL/GDL interface forms,
which works as an ideal non polarizable catalyst layer with fast oxygen
transport and large ORR rate. The concerted
action of both effects leads to dramatic lowering of the CCL transport
impedance.

Further increase in $k$ changes the sign of the phase angle at
the membrane interface, meaning formation of inductive loop in the Nyquist spectrum.
In this state, the static resistivity drops below the charge--transfer
value  under constant oxygen supply $R_{ct} = b/j$;
this regime deserves special consideration and it is not discussed here.
The limiting value $k_{\lim}$ at which $\tR = \tR_{ct} = 1 / \tj_0$ is
\begin{equation}
   k_{\lim} = \tc_1^0\bigl(\phi\coth(\phi) - 1\bigr)
   \label{eq:klim}
\end{equation}
The states with $k > k_{\lim}$ could be  unstable, which needs further studies.

\begin{table}
    \small
\begin{center}
    \begin{tabular}{|l|c|}
        \hline
        Tafel slope $b$, V                   & 0.03 \\
        Double layer capacitance $\Cdl$, F~cm$^{-3}$  &  20 \\
        Exchange current density $i_*$, A~cm$^{-3}$  &  $10^{-3}$  \\
        Oxygen diffusion coefficient         & \\
        in the CCL, $\Dox$,  cm$^2$~s$^{-1}$ &  $1\cdot 10^{-5}$ \\
        Catalyst layer thickness $\lcat$, cm & $10\cdot 10^{-4}$       \\
        \hline
        Cell temperature $T$, K                        & 273 + 80     \\
        Cathode pressure, bar                    & 1.5 \\
        Cathode relative humidity, \%            & 50 \\
        Cell current density $j_0$, A~cm$^{-2}$  &  $0.1$   \\
        \hline
    \end{tabular}
\end{center}
    \caption{The cell parameters used in calculations. Parameter $\Dox$ is taken to be small
    to emphasize the effect of oxygen transport in the CCL.
    }
    \label{tab:parms}
\end{table}

From practical perspective, in a real fuel cell it is not feasible to directly
perturb the oxygen concentration at the CCL/GDL interface
in--phase with the overpotential perturbation. However,
the  experiments of Kim \etal~\cite{Kim_08b} and Hwang \etal~\cite{Hwang_10}
suggest that the effect could be achieved by perturbing air flow velocity in the channel.
Such a perturbation could cause in--phase oscillations of the oxygen
concentration and overpotential, which greatly improves the transient cell performance.


\begin{thebibliography}{13}
\providecommand{\natexlab}[1]{#1}
\providecommand{\url}[1]{\texttt{#1}}
\expandafter\ifx\csname urlstyle\endcsname\relax
  \providecommand{\doi}[1]{doi: #1}\else
  \providecommand{\doi}{doi: \begingroup \urlstyle{rm}\Url}\fi

\bibitem[Lasia(2014)]{Lasia_book_14}
A.~Lasia.
\newblock \emph{Electrochemical Impedance Spectroscopy and its Applications}.
\newblock Springer, New York, 2014.

\bibitem[Engebretsen et~al.(2017)Engebretsen, Mason, Shearing, Hinds, and
  Brett]{Brett_17}
E.~Engebretsen, T.~J. Mason, P.~R. Shearing, G.~Hinds, and D.~J.~L. Brett.
\newblock Electrochemical pressure impedance spectroscopy applied to the study
  of polymer electrolyte fuel cells.
\newblock \emph{Electrochem.\ Comm.}, 75:\penalty0 60--63, 2017.
\newblock \doi{10.1016/j.elecom.2016.12.014}.

\bibitem[Shirsath et~al.(2020)Shirsath, Rael, Bonnet, Schiffer, Bessler, and
  Lapicque]{Bessler_20}
A.~V. Shirsath, S.~Rael, C.~Bonnet, L.~Schiffer, W.~Bessler, and F.~Lapicque.
\newblock Electrochemical pressure impedance spectroscopy for investigation of
  mass transfer in polymer electrolyte membrane fuel cells.
\newblock \emph{Current Opinion in Electrochem.}, 20:\penalty0 82--87, 2020.
\newblock \doi{10.1016/j.coelec.2020.04.017}.

\bibitem[Schiffer et~al.(2021)Schiffer, Shirsath, Ra{\"e}l, Lapicque, and
  Bessler]{Bessler_22}
L.~Schiffer, A.~V. Shirsath, S.~Ra{\"e}l, F.~Lapicque, and W.~G. Bessler.
\newblock Electrochemical pressure impedance spectroscopy for polymer
  electrolyte membrane fuel cells: {A} combined modeling and experimental
  analysis.
\newblock \emph{J.\ Electrochem.\ Soc.}, 169:\penalty0 034503, 2021.
\newblock \doi{10.1149/1945-7111/ac55cd}.

\bibitem[Zhang et~al.(2022)Zhang, Homayouni, Gates, Eikerling, and
  Niroumand]{Zhang_22}
Q.~Zhang, H.~Homayouni, B.~D. Gates, M.~Eikerling, and A.~M. Niroumand.
\newblock Electrochemical pressure impedance spectroscopy for polymer
  electrolyte fuel cells via back-pressure control.
\newblock \emph{J.\ Electrochem.\ Soc.}, 169:\penalty0 044510, 2022.
\newblock \doi{10.1149/1945-7111/ac6326}.

\bibitem[Sorrentino et~al.(2017)Sorrentino, Vidakovic-Koch, Hanke-Rauschenbach,
  and Sundmacher]{Sorrentino_17}
A.~Sorrentino, T.~Vidakovic-Koch, R.~Hanke-Rauschenbach, and K.~Sundmacher.
\newblock Concentration--alternating frequency response: {A} new method for
  studying polymer electrolyte membrane fuel cell dynamics.
\newblock \emph{Electrochim.\ Acta}, 243:\penalty0 53--64, 2017.
\newblock \doi{10.1016/j.electacta.2017.04.150}.

\bibitem[Sorrentino et~al.(2019)Sorrentino, Vidakovic-Koch, and
  Sundmacher]{Sorrentino_19}
A.~Sorrentino, T.~Vidakovic-Koch, and K.~Sundmacher.
\newblock Studying mass transport dynamics in polymer electrolyte membrane fuel
  cells using concentration--alternating frequency response analysis.
\newblock \emph{J.\ Power Sources}, 412:\penalty0 331--335, 2019.
\newblock \doi{10.1016/j.jpowsour.2018.11.065}.

\bibitem[Sorrentino et~al.(2020)Sorrentino, Sundmacher, and
  Vidakovic-Koch]{Sorrentino_20}
A.~Sorrentino, K.~Sundmacher, and T.~Vidakovic-Koch.
\newblock Polymer electrolyte fuel cell degradation mechanisms and their
  diagnosis by frequency response analysis methods: A review.
\newblock \emph{Energies}, 13:\penalty0 5825, 2020.
\newblock \doi{10.3390/en13215825}.

\bibitem[Kim et~al.(2008)Kim, Han, Kim, and Rhee]{Kim_08b}
Y.~H. Kim, H.~S. Han, S.~Y. Kim, and G.~H. Rhee.
\newblock Influence of cathode flow pulsation on performance of
  proton--exchange membrane fuel cell.
\newblock \emph{J.\ Power Sources}, 185:\penalty0 112--117, 2008.
\newblock \doi{10.1016/j.jpowsour.2008.06.069}.

\bibitem[Hwang et~al.(2010)Hwang, Lee, Choi, Kim, Cho, Joonho, Kim, Jang, Kim,
  and Cha]{Hwang_10}
Y.-S. Hwang, D.-Y. Lee, J.~W. Choi, S.-Y. Kim, S.~H. Cho, P.~Joonho, M.~S. Kim,
  J.~H. Jang, S.~H. Kim, and S.-W. Cha.
\newblock Enhanced diffusion in polymer electrolyte membrane fuel cells using
  oscillating flow.
\newblock \emph{Int.\ J.\ Hydrogen Energy}, 35:\penalty0 3676--3683, 2010.
\newblock \doi{10.1016/j.ijhydene.2010.01.064}.

\bibitem[Kulikovsky(2020)]{Kulikovsky_20arXiv}
Andrei Kulikovsky.
\newblock Performance of a {PEM} fuel cell with oscillating air flow velocity:
  {A} modeling study based on cell impedance.
\newblock \emph{arXiv}, 2020.
\newblock \doi{10.48550/arXiv.2008.07101}.

\bibitem[Kulikovsky(2021)]{Kulikovsky_21a}
Andrei Kulikovsky.
\newblock Performance of a {PEM} fuel cell with oscillating air flow velocity:
  {A} modeling study based on cell impedance.
\newblock \emph{eTrans.}, 5:\penalty0 100104, 2021.
\newblock \doi{10.1016/j.etran.2021.100104}.

\bibitem[Kulikovsky(2016)]{Kulikovsky_16a}
A.~A. Kulikovsky.
\newblock A simple physics--based equation for low--current impedance of a
  {PEM} fuel cell cathode.
\newblock \emph{Electrochim.\ Acta}, 196:\penalty0 231--235, 2016.
\newblock \doi{10.1016/j.electacta.2016.02.150}.

\end{thebibliography}

\small

\section*{Nomenclature}

\begin{tabular}{ll}
    $\tilde{}$   &  Marks dimensionless variables                             \\
    $b$          &  ORR Tafel slope, V                                        \\
    $\Cdl$       &  Double layer volumetric capacitance, F~cm$^{-3}$          \\
    $c$          &  Oxygen molar concentration, mol~cm$^{-3}$                 \\
    $\cref$      &  Reference oxygen concentration, mol~cm$^{-3}$             \\
    $\Dox$       &  Oxygen diffusion coefficient in the CCL, cm$^2$~s$^{-1}$ \\
    $F$          &  Faraday constant, C~mol$^{-1}$                            \\
    $f$          &  Frequency, Hz                                    \\
    $G$          &  Dimensionless concentration admittance, Eq.\eqref{eq:YG}  \\
    $j$          &  Local proton current density, A~cm$^{-2}$                   \\
    $\ri$        &  Imaginary unit                                            \\
    $i_*$        &  ORR volumetric exchange current density, A~cm$^{-3}$          \\
    $k$          &  Concentration amplitude factor, Eq.\eqref{eq:atcx}            \\
    $\lcat$      &  Catalyst layer thickness, cm                              \\
    $R$          &  Static resistivity, $\Omega$~cm$^2$                       \\
    $t$          &  Time, s                                                   \\
    $x$          &  Coordinate through the cell, cm \\
    $Y$          &  Dimensionless electric  admittance, Eq.\eqref{eq:YG}  \\
    $Z$          &  Impedance,  $\Omega$~cm$^2$                      \\ [1em]
\end{tabular}


{\bf Subscripts:\\}

\begin{tabular}{ll}
    $RC$     & parallel $RC$--circuit impedance \\
    $W$      & Warburg finite--length impedance \\
    $0$      & membrane/CCL interface       \\
    $1$      & CCL/GDL interface  \\[1em]
\end{tabular}

{\bf Superscripts:\\}

\begin{tabular}{ll}
    $0$      & Steady--state value            \\
    $1$      & Small perturbation  amplitude \\[1em]
\end{tabular}


{\bf Greek:\\}

\begin{tabular}{ll}
    $\eta$              &  ORR overpotential, positive by convention, V   \\
    $\eta_1^1$          &  Amplitude of applied overpotential perturbation, V   \\
    $\mu$               &  Dimensionless parameter, Eq.\eqref{eq:mu}      \\
    $\phi$              &  Dimensionless parameter, Eq.\eqref{eq:tRccl}    \\
    $\psi$              &  Dimensionless parameter, Eq.\eqref{eq:Gsol}    \\
    $\omega$            &  Angular frequency of the AC signal, s$^{-1}$
\end{tabular}

\newpage

\end{document}